\documentclass[a4paper,12pt]{article}
\usepackage{amsbsy}
\usepackage[fleqn]{amsmath}
\usepackage{amssymb}
\usepackage{array}
\usepackage{booktabs}
\usepackage{xcolor}
\usepackage[colorlinks, linkcolor=blue, anchorcolor=blue, citecolor=blue]{hyperref}
\usepackage{mathrsfs}
\usepackage{tabularx}
\usepackage{latexsym}
\usepackage{textcomp}
\usepackage{pstricks}
\usepackage{colortbl}
\usepackage{multirow}
\usepackage{fancyhdr}
\usepackage{bm}
\usepackage{booktabs}
\usepackage{subfigure}
\usepackage{caption}
\usepackage{soul}
\usepackage{cleveref}
\usepackage{cite}
\usepackage{lineno}
\usepackage{ulem}
\usepackage{graphicx}
\usepackage{epstopdf}
\usepackage{indentfirst}
\usepackage{setspace}
\usepackage[latin1]{inputenc}
\usepackage{titlesec}
\usepackage{tabularx} % for 'tabularx' env. and 'X' col. type
\usepackage{ragged2e} % for \RaggedRight macro
\usepackage{booktabs} % for \toprule, \midrule etc macros
\usepackage{qcircuit} % Quantum circuits
\usepackage{braket}   %braket sign
\usepackage{svg}      % Insert svg: \includesvg{filename}
\usepackage{amsmath}

\usepackage{algorithm}
\usepackage{algpseudocode}
\usepackage{mdframed}
\usepackage{xcolor}

\titleformat*{\section}{\large\bf}
\titleformat*{\subsection}{\normalsize\bf}
\crefname{section}{Section}{Sections}
\crefname{equation}{Eq.}{Eqs.}
\crefname{figure}{Figure}{Figs.}
\crefname{table}{Table}{Tables}
\crefname{algorithm}{Algorithm}{Algorithms}
\begin{document}

% \linenumbers%Used for the line numbers

%
%
%%%%%%%%%%%%%%%%%%%%%%%%%%%%%%%%%%%%%%%%%%%%%%%%%%%%
%%%                     TITLE                    %%%
%%%%%%%%%%%%%%%%%%%%%%%%%%%%%%%%%%%%%%%%%%%%%%%%%%%%
\begin{center}
\Large\textbf{Quantum Computing Enhanced Distance-Minimizing Data-Driven Computational Mechanics}
\end{center}
\large{
\begin{center}
\textbf{Yongchun Xu}$^{a}$, \textbf{Jie Yang$^{a}$}, \textbf{Zengtao Kuang$^{a}$}, \textbf{Qun Huang$^{a}$},\\ \textbf{Wei Huang$^{a}$}, \textbf{Heng Hu}$^{a,}$\footnote{Corresponding author. E-mail address: huheng@whu.edu.cn.}$^{}$
\end{center}
}
\small{
\begin{center}
$^a$School of Civil Engineering, Wuhan University, 8 South Road of East Lake, Wuchang, 430072 Wuhan, PR China\\
\end{center}
}

%
%
%%%%%%%%%%%%%%%%%%%%%%%%%%%%%%%%%%%%%%%%%%%%%%%%%%%%
%%%                     ABSTRACT                 %%%
%%%%%%%%%%%%%%%%%%%%%%%%%%%%%%%%%%%%%%%%%%%%%%%%%%%%
\begin{flushleft}
\large\textbf{Abstract}
\end{flushleft}
\indent\indent The distance-minimizing data-driven computational mechanics has great potential in engineering applications by eliminating material modeling error and uncertainty. In this computational framework, the solution-seeking procedure relies on minimizing the distance between the constitutive database and the conservation law. However, the distance calculation is time-consuming and often takes up most of the computational time in the case of a huge database. In this paper, we show how to use quantum computing to enhance data-driven computational mechanics by exponentially reducing the computational complexity of distance calculation. The proposed method is not only validated on the quantum computer simulator Qiskit, but also on the real quantum computer from OriginQ. We believe that this work represents a promising step towards integrating quantum computing into data-driven computational mechanics.

\begin{flushleft}
\justifying\textbf{Keywords:} Data-driven computational mechanics; Quantum computing; Distance calculation; Swap test; Nearest-neighbor search.
\end{flushleft}

%
%
%%%%%%%%%%%%%%%%%%%%%%%%%%%%%%%%%%%%%%%%%%%%%%%%%%%%
%%%                INTRODUCTION                  %%%
%%%%%%%%%%%%%%%%%%%%%%%%%%%%%%%%%%%%%%%%%%%%%%%%%%%%
\section{Introduction}

The distance-minimizing data-driven computational mechanics (DDCM) \cite{Ortiz2016} is a new computing paradigm for solving boundary value problems (BVP) in science and engineering. During the past few years, it has been rapidly extended to dynamics \cite{Ortiz2018dynamics}, large deformations \cite{kuang2023data,kuang2023data2}, inelasticity \cite{eggersmann2019model}, and fracture mechanics \cite{carrara2020data,carrara2021data,huang2021data}. Meanwhile, in combination with offline computational homogenization \cite{xu2020Data,karapiperis2021data}, the DDCM has accelerated the online multi-scale analysis of composite materials and structures \cite{yang2019structural,yan2022data,hui2022data,bai2022data}. It is worth noting that, with material data from different physical fields, the DDCM has also been employed for analyzing magnetic problems \cite{de2020magnetic} and coupled electro-mechanical problems \cite{marenic2022data}. Recently, a unified functional based coupling framework that integrates the DDCM into the traditional finite element method has been successfully applied in local refinement problems \cite{yang2022investigation,wattel2023mesh,yang2023unified}. Compared to classic computational mechanics, DDCM does not rely on constitutive models, which is beneficial from two aspects: (1) the complex material modeling process that takes huge intellectual and time costs is avoided; (2) errors and uncertainties generated during the fitting process might also be eliminated.

Although DDCM has great potential in scientific computing, it currently suffers from computational efficiency issues. One of the key issues is the efficiency of data searching, especially when dealing with large-scale databases. To be more specific,  assume that a data search is conducted in a database with $N$ data, the dimension of data is $D$, which means each data contains $D$ variables. In one nearest neighbor search, the computational complexity on a classical computer requires executing $N$ distance calculations, each with a cost of $O(D)$, resulting in the total computational complexity $O(ND)$. Therefore, the efficiency of distance calculation can be improved in two ways: the first one is to reduce the number of distance calculations $N$, and the second one is to directly reduce the distance calculation cost $O(D)$. The former has been widely studied and is achieved through different approaches \cite{korzeniowski2021multi,kdtree2,bahmani2021kd}. For instance, $k$-d tree \cite{friedman1977algorithm} provides an appropriate data structure to optimize the number of queries in the database, resulting in the computational complexity from $O(ND)$ to approximately $O(\log(N)D)$. For the latter, however, the calculation efficiency is difficult to improve, because the data dimension reduction is an inherent challenge.

The emergence of the quantum computer offers the possibility to overcome certain limitations of the classical computer. It facilitates the reduction of data dimension when processing the distance calculation. Based on the swap test quantum circuit proposed by Buhrman et al. \cite{buhrman2001quantum}, Seth Lloyd et al. \cite{lloyd2013quantum} develop a quantum algorithm subroutine to efficiently estimate the distance between two data, which reduces the computational complexity from $O(D)$ to $O(\log D)$. To encode the data into the quantum computer, the quantum random access memory (qRAM) \cite{giovannetti2008quantum} is used, which is the quantum counterpart to the classical random access memory (RAM). This subroutine has been adopted in a variety of quantum machine learning algorithms, such as quantum SVM \cite{rebentrost2014quantum}, q-means \cite{kerenidis2019q}, and quantum KNN \cite{wiebe2015quantum}. Therefore, quantum computing can clearly be exploited in the distance calculation during the data searching of DDCM.

In this paper, we propose a quantum computing enhanced data-driven computational framework, referred to as qDD. It aims to replace the distance calculation of the nearest-neighbor search in classical computing with the quantum algorithm. In this way, the computational complexity of the distance estimation reduces from $O(ND)$ to $O(N \log D)$. To demonstrate the convergence and accuracy of qDD, several numerical examples are investigated. The calculations are carried out on both the quantum computer simulator Qiskit from IBM \cite{Qiskit} and the real quantum computer WuYuan from OriginQ \cite{Origin}.

The layout of this paper is as follows. In \cref{Methodology},  the main idea of data-driven computing is recalled first, then the quantum algorithm for distance estimation along with its error analysis and error reduction are introduced.  In \cref{Validation}, the validation of qDD using the quantum computer simulator Qiskit is presented. In \cref{real}, an experiment on a real quantum computer WuYuan from OriginQ is presented. In \cref{plate}, a two-dimensional numerical test example using the simulator is shown. Conclusions and discussions are outlined in \cref{Discussion}.

%
%
%%%%%%%%%%%%%%%%%%%%%%%%%%%%%%%%%%%%%%%%%%%%%%%%%%%%
%%%                 Methodology                  %%%
%%%%%%%%%%%%%%%%%%%%%%%%%%%%%%%%%%%%%%%%%%%%%%%%%%%%
\section{Methodology}\label{Methodology}

In this section, the basic formulations of data-driven computing are briefly recalled in \cref{DD-solver}, and the quantum algorithm for distance estimation of two stress-strain data is presented in \cref{Distance}. The error analysis and the error reduction of the quantum algorithm are proposed in \cref{Adaptive}. Finally, we summarize the framework of qDD and discuss its computational complexity in \cref{complexity}.

%%%%%%%%%%%%%%%%%%%%%%%%%%%%%%%%%%%%%%%%%%%%%%%%%%%%%%%%%%%%%%%%%%%%%%%%%%%%%%%%%
\subsection{Overview of data-driven computing}\label{DD-solver}

We consider a finite element discretization of an elastic solid, the material behavior of which is characterized by a set of stress-strain data in a material database $\mathscr{D}$. The data-driven computing seeks to assign to each integration point $e$ the optimal stress-strain data $\bm{z}^{*}_e=(\bm{\epsilon}^{*}_{e}, \bm{\sigma}^{*}_{e})\in\mathscr{D}$, which is closest to the admissible stress-strain state $\bm{z}_e=(\bm{\epsilon}_{e}, \bm{\sigma}_{e})$ that satisfies equilibrium and compatibility. Such a constrained minimization problem can be expressed with a distance-based functional as follows:
\begin{equation}\label{Eq01}
	\Pi=\sum_{e=1}^{m}\frac{1}{2}{w_e}\mathscr{F}_e(\bm{z}_e,\bm{z}_e^*)-\bm{\eta}^\text{T}\left(\sum_{e=1}^{m}{w}_e\textbf{B}_e\bm{\sigma}_e-\bm{f}\right)
\end{equation}
with the distance term reads:
\begin{equation}\label{Eq02}
	\mathscr{F}_e(\bm{z}_e,\bm{z}_e^*) = (\bm{\epsilon}_e-\bm{\epsilon}_e^*)^{\text{T}}{\mathbb{C}}(\bm{\epsilon}_e-\bm{\epsilon}_e^*) + (\bm{\sigma}_e-\bm{\sigma}_e^*)^{\text{T}}{\mathbb{C}}^{-1}(\bm{\sigma}_e-\bm{\sigma}_e^*)
\end{equation}
where $\bm{\eta}$ is a vector of Lagrange multipliers enforcing the equilibrium constraints, $\textbf{B}_e$ is a matrix of interpolation functions, $\bm{f}$ represents the concentrated nodal force, $\mathbb{C}$ is a user-defined symmetric matrix to scale stress and strain to a similar magnitude, the superscript $\text{T}$, $w_{e}$ and $m$ respectively denote the transpose symbol, the integration weight and the total number of integration points.

Considering the compatibility constraints $\bm{\epsilon}_{e}=\textbf{B}_{e}\bm{u}$, one takes all possible variations of \cref{Eq02} and gets the following linear equations:
\begin{subequations}\label{Eq03}
\begin{align}
& \delta \bm{u} \quad \Rightarrow \quad\sum_{e=1}^{m} w_{e} \textbf{B}_{e}^\text{T} \mathbb{C}\textbf{B}_{e}\bm{u}=\sum_{e=1}^{m} w_{e} \textbf{B}_{e}^\text{T} \mathbb{C}\bm{\epsilon}_{e}^{*}\label{Eq03_1}\\
& \delta \bm{\sigma}_{e} \quad \Rightarrow \quad \bm{\sigma}_{e}=\bm{\sigma}_{e}^{*}+\mathbb{C}\textbf{B}_{e}\bm{\eta}\label{Eq03_2}\\
& \delta \bm{\eta}\quad \Rightarrow \quad\sum_{e=1}^m w_e \textbf{B}_{e}^\text{T} \bm{\sigma}_e=\bm{f}\label{Eq03_3}
\end{align}
\end{subequations}
By substituting \cref{Eq03_2} into \cref{Eq03_3}, the linear system can be reduced as follows:
\begin{subequations}\label{DD_linear}
\begin{align}
& \sum_{e=1}^{m} w_{e} \textbf{B}_{e}^\text{T} \mathbb{C}\textbf{B}_{e}\bm{u}=\sum_{e=1}^{m} w_{e} \textbf{B}_{e}^\text{T} \mathbb{C}\bm{\epsilon}_{e}^{*} \\
& \sum_{e=1}^m w_e \textbf{B}_{e}^\text{T} \mathbb{C} \textbf{B}_{e} \bm{\eta}=\bm{f}-\sum_{e=1}^m w_e \textbf{B}_{e}^\text{T} \bm{\sigma}_e^*
\end{align}
\end{subequations}

The resulting linear system can be solved iteratively, in which randomly selected data $\bm{z}_e^*$ from the database $\mathscr{D}$ is assigned to each integration point to initialize the data-driven computing. This iteration process consists of a mapping operation from $\bm{z}_e^*$ to $\bm{z}_e$ by solving \cref{DD_linear} and a reverse mapping from $\bm{z}_e$ to $\bm{z}_e^*$ via nearest-neighbor search that satisfies the following condition:
\begin{equation}
	\mathscr{F}_e(\bm{z}_e,\bm{z}_e^*)
	\leq
	\mathscr{F}_e(\bm{z}_e,\bm{z}'_e),\quad\forall \bm{z}'_e \in \mathscr{D}
\end{equation}
These two steps are performed alternately until the distance between $\bm{z}_e$ to $\bm{z}_e^*$ is minimized for each integration point. It is worth noting that massive distance calculations during the nearest-neighbor search lead to extremely high computational complexity when faced with high-dimensional and high-density databases. Quantum algorithm for distance estimation makes it possible to reduce the computational complexity, which will be introduced in the following.

%%%%%%%%%%%%%%%%%%%%%%%%%%%%%%%%%%%%%%%%%%%%%%%%%%%%%%%%%%%%%%%%%%%%%%%%%%%%%%%%%
\subsection{Distance estimation via a quantum algorithm}\label{Distance}

We consider one integration point $e$ and study the nearest-neighbor search for the corresponding admissible data $\bm{z}_e$. To find its nearest material data $\bm{z}_e^*$ in the database $\mathscr{D}$, we need to calculate the distances between $\bm{z}_e$ and all ${\bm{z}'_e\in\mathscr{D}}$ and find the minimal one. For the sake of simplicity, we use $d$ to denote the distance $\mathscr{F}_e(\bm{z}_e,\bm{z}_e')$ in \cref{Eq02}, 
it is expressed in the form of the squared of the Euclidean distance:
\begin{equation}\label{dis2}
d=\left|\bm{X}_{e}-\bm{X}'_{e}\right|^2
\end{equation}
where $\bm{X}_{e}$ and $\bm{X}'_{e}$ are the $D$-dimension vectors of the scaled admissible data $\bm{z}_e$ and the scaled material data $\bm{z}'_e$, respectively.
In addition, the scaled optimal data $\bm{z}_e^*$ can be denoted as $\bm{X}^{*}_{e}$.

To apply the quantum algorithm for the distance estimation, the information needs to be represented  with quantum states. Here, the amplitude encoding is used to represent the information with the coefficients of the quantum basis states, which can represent a $D$-dimensional vector with only $\log_{2} D$ qubits. Note that the square of the coefficients, which represents the probabilities associated with measuring the corresponding basis states, must collectively sum to 1 \cite{Nielsen}. Therefore, the normalized vectors $\bm{X}_{e}$ and $\bm{X}'_{e}$ can be represented as:
\begin{subequations}\label{states}
	\begin{align}
	& |\bm{X}_{e}\rangle=\frac{1}{\left| \bm{X}_{e} \right|}\sum_{i=1}^{D} x_i |i\rangle\\
	& |\bm{X}'_{e}\rangle=\frac{1}{\left| \bm{X}'_{e} \right|}\sum_{i=1}^{D} x'_i |i\rangle
	\end{align}
\end{subequations}
In the above equations, the Dirac notation $|\cdot\rangle$ is employed to represent a quantum state.
The components of $\bm{X}_{e}$ and $\bm{X}'_{e}$ are denoted as $x_i$ and $x'_i$, respectively. The symbol $|i\rangle$ denotes a state in the computational basis.
Furthermore, to transfer information from classical computers to quantum computers, a quantum device called qRAM \cite{giovannetti2008quantum} is used. For example, to encode $|\bm{X}_{e}\rangle$ into a quantum computer, only $O(\log D)$ operations are required to access the data \cite{giovannetti2008quantum}. 
With the help of qRAM, the following two quantum states are encoded into the quantum computer:
\begin{subequations}\label{states2}
\begin{align}
& |\phi\rangle=\frac{1}{\sqrt{Z}}(\left| \bm{X}_{e} \right||0\rangle-\left| \bm{X}'_{e} \right||1\rangle)\\
& |\psi\rangle=\frac{1}{\sqrt{2}}(|0\rangle|\bm{X}_{e}\rangle+|1\rangle|\bm{X}'_{e}\rangle)
\end{align}
\end{subequations}
where $Z=\left| \bm{X}_{e} \right|^2+\left| \bm{X}'_{e} \right|^2$,  $|\phi\rangle$ contains the norms of the two data and requires only 1 qubit, $|\psi\rangle$ contains the two normalized data and needs $\log_{2} 2D$ qubits. 
$Z$ and components of $|\phi\rangle$ and $|\psi\rangle$ need to be pre-computed before running the quantum algorithm. Once the inner product of $|\phi\rangle$ and $|\psi\rangle$ is calculated, the distance $d$ is obtained through the following relation:
\begin{equation}\label{d}
|\langle \phi| \psi \rangle|^2 = \frac{\left|\bm{X}_{e}-\bm{X}'_{e}\right|^2}{2Z} =\frac{d}{2Z}
\end{equation}

\begin{figure}[t]
\centering
\includegraphics[width=8cm]{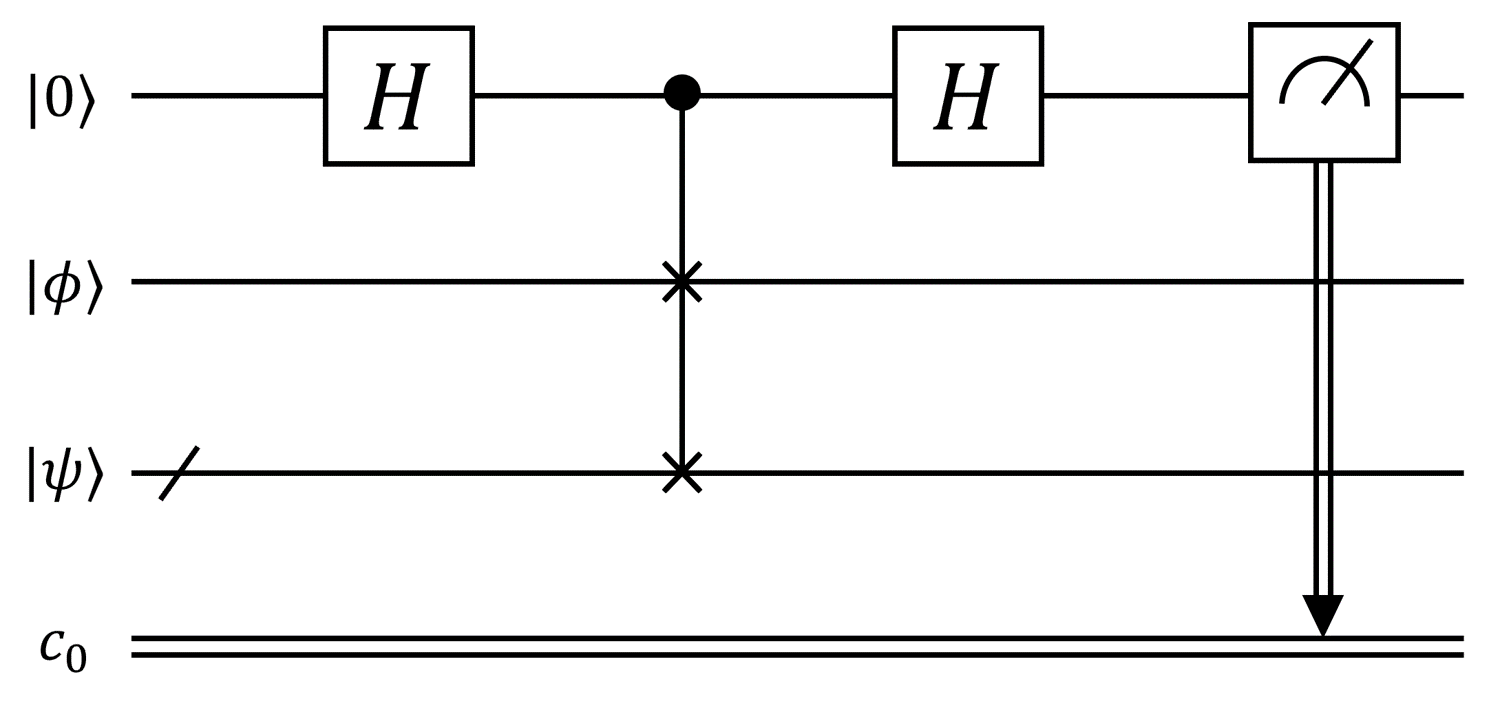}
\caption{Quantum circuit of the swap test.}
\label{swap_fig}
\end{figure}

To obtain the inner product $|\langle \phi| \psi \rangle|^2$, the swap test is applied \cite{buhrman2001quantum}. The corresponding quantum circuit is shown in \cref{swap_fig}, 
where the quantum states $|\phi\rangle$ and $|\psi\rangle$ are encoded at the beginning of the circuit, as well as an ancilla qubit in $|0\rangle$ and a classical bit $c_0$. Then three quantum gates, including two Hadamard gates and one controlled-SWAP gate, are utilized to manipulate the quantum information.

At the end of the circuit, the measurement of the ancilla qubit is performed, and the result is saved into the classical bit $c_0$. The state of $c_0$ is 0 or 1, corresponding to the quantum state of the ancilla qubit measured to be in state $|0\rangle$ or $|1\rangle$, and the possibility of being in state $|0\rangle$ is:
\begin{equation}\label{p}
\begin{split}
p = \frac{1}{2} + \frac{1}{2}{| {\left\langle \phi  | \psi  \right\rangle } |^2}
\end{split}
\end{equation}
Substitute \cref{d} into \cref{p}, we get the desired distance:
\begin{equation}\label{dis_final}
d=\left|\bm{X}_{e}-\bm{X}'_{e}\right|^2=4Z(p-\frac{1}{2})
\end{equation}
The key procedure of the distance estimation using this quantum algorithm is to obtain an estimated value $\bar{p}$ for $p$, then the estimated distance $\bar{d}$ for $d$ is calculated using \cref{dis_final}. The computational complexity of this quantum algorithm is $O(\log D)$, which is an exponential speed-up compared to classical computing. The detailed complexity analysis is presented in \cref{complexity}.

Due to the inherently probabilistic nature of quantum systems, extracting meaningful information from a quantum computer necessitates multiple runs of the quantum algorithm, with information obtained from the distribution of the quantum state. The estimated distance $\bar{d}$ is calculated by estimating $\bar{p}$, which may not perfectly match the true probability $p$, leading to computational errors in $\bar{d}$. This error makes qDD hard to converge if we directly implement the quantum algorithm in the nearest-neighbor search. In the next section, we present the estimation of $p$, and the analysis of the error of the quantum algorithm. Furthermore, a strategy will be proposed to reduce the error.

%%%%%%%%%%%%%%%%%%%%%%%%%%%%%%%%%%%%%%%%%%%%%%%%%%%%%%%%%%%%%%%%%%%%%%%%%%%%%%%%%
\subsection{Error analysis and reduction}\label{Adaptive}

 In this section, we analyze the error of the quantum algorithm, and an adaptive strategy is proposed to reduce the error to improve the convergence of qDD. Note that physical quantum computers are sensitive, and their results might be affected by various sources, such as environmental noise and imperfect gate implementations. In this work, we assume the implementation of qDD on a fault-tolerant quantum computer, hardware errors are thus not taken into consideration.

\subsubsection{Error analysis}\label{Error_A}

\begin{figure}[t]
\centering
\includegraphics[width=10cm]{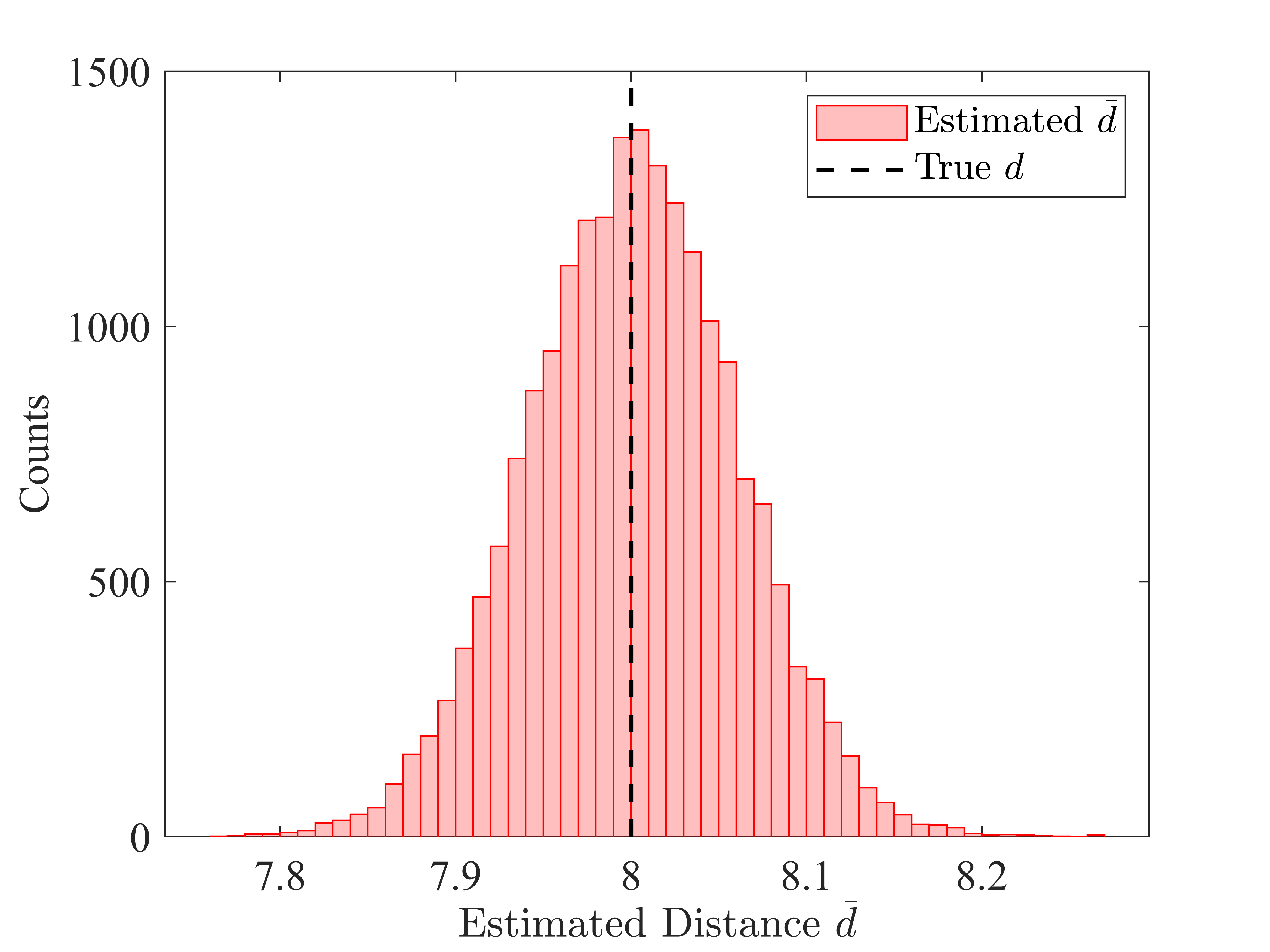}
\caption{The distribution of the estimated distance $\bar{d}$ between two data $\bm{X}_{e}=[0.5,-1.5]$ and $\bm{X}'_{e}=[-1.5,0.5]$.}
\label{dis_example}
\end{figure}

A straightforward illustration is provided in \cref{dis_example}, to enlighten the purpose of the error analysis. 
% To illustrate the purpose of error analysis, we provide a straightforward illustration in \cref{dis_example}. 
This presents the distribution of the estimated distance $\bar{d}$ between two data, $\bm{X}_{e}=[0.5,-1.5]$ and $\bm{X}'_{e}=[-1.5,0.5]$. The distribution comprises 20000 samples of $\bar{d}$, which were obtained using the quantum computer simulator Qiskit. The true distance $d$ is 8, marked with the black dashed line. It is evident that the estimated $\bar{d}$ values cluster around the true distance $d$, indicating an estimation error. The primary objective of this subsection is to determine the root-mean-square error (RMSE) of the estimated $\bar{d}$, which  necessitates the statistical analysis of the quantum algorithm.

We run the quantum algorithm multiple times to get an estimated $\bar{p}$. Let us denote the number of repetitions as $n_s$, and we have $n_s$ measurement results. As is introduced in \cref{Distance}, one measurement corresponds to a classical bit $c_0$ in state 0 with probability $p$ and in state 1 with  probability $1-p$. 
%Acoording to \cref{p}, the probability of obtaining 0 is $p=\frac{1}{2} + \frac{1}{2}{\left| {\left\langle \phi  | \psi  \right\rangle } \right|^2}$. 
%
If we denote the number of 0s in all measurement results as $V$, then it follows a binomial distribution denoted by $V \sim B(n_s,p)$. The estimated $\bar{p}$ can be derived from the maximum-likelihood estimation of $p$:
\begin{equation}\label{estimated_p}
\bar p = \frac{V}{{{n_s}}}
\end{equation}
Therefore, the estimated $\bar{d}$ is calculated using \cref{dis_final,estimated_p}:
\begin{equation}\label{estimated_d}
{{\bar d}} = Z(4\frac{V}{{{n_s}}} - 2)
\end{equation}

Now we analyze the error of the estimated $\bar{d}$. Given that $V \sim B(n_s,p)$, it's well-known that the mean of $V$ is ${n_s}p$, and the variance is ${n_s}p(1 - p)$. 
Thus, we can derive the mean of the estimated $\bar{d}$:
\begin{equation}\label{Mean_d}
\mu_{\bar d} = Z(4\frac{{n_s}p}{{{n_s}}} - 2)=d
\end{equation}
which is equal to $d$, indicating  that the estimation of the distance is unbiased. Then, we can derive the RMSE of $\bar{d}$ to represent the estimation error:
\begin{equation}\label{Var_d}
\epsilon_{\bar d} = \sqrt{\frac{4Z^2-d^2}{{{n_s}}}}
\end{equation}
It can be proved that:
\begin{equation}\label{Var_d3}
0 \leq \epsilon_{\bar d} \leq \sqrt{\frac{4Z^2}{{{n_s}}}}
\end{equation}
This indicates that the error of the distance estimation $\epsilon_{\bar{d}}$ can be reduced in two ways: (1) increasing the number of measurements $n_s$; (2) reducing $Z=\left| \bm{X}_e \right|^2+\left| \bm{X}'_e \right|^2$. 
For the first way, to get an estimation result to accuracy $\epsilon_{\bar d}$, we need to run the quantum algorithm $O({Z^2}/{{\epsilon_{\bar d}^2}})$ times. 
The error has limit zero as $n_s$ tends to infinity, but it leads to large computational costs.
%When $n_s$ is large enough, the distance estimation error approaches null. However, it leads to large computational costs. 
In this paper, we prefer to use the second way (see details in the next subsection).

\begin{figure}[t]
\centering
\includegraphics[width=13cm]{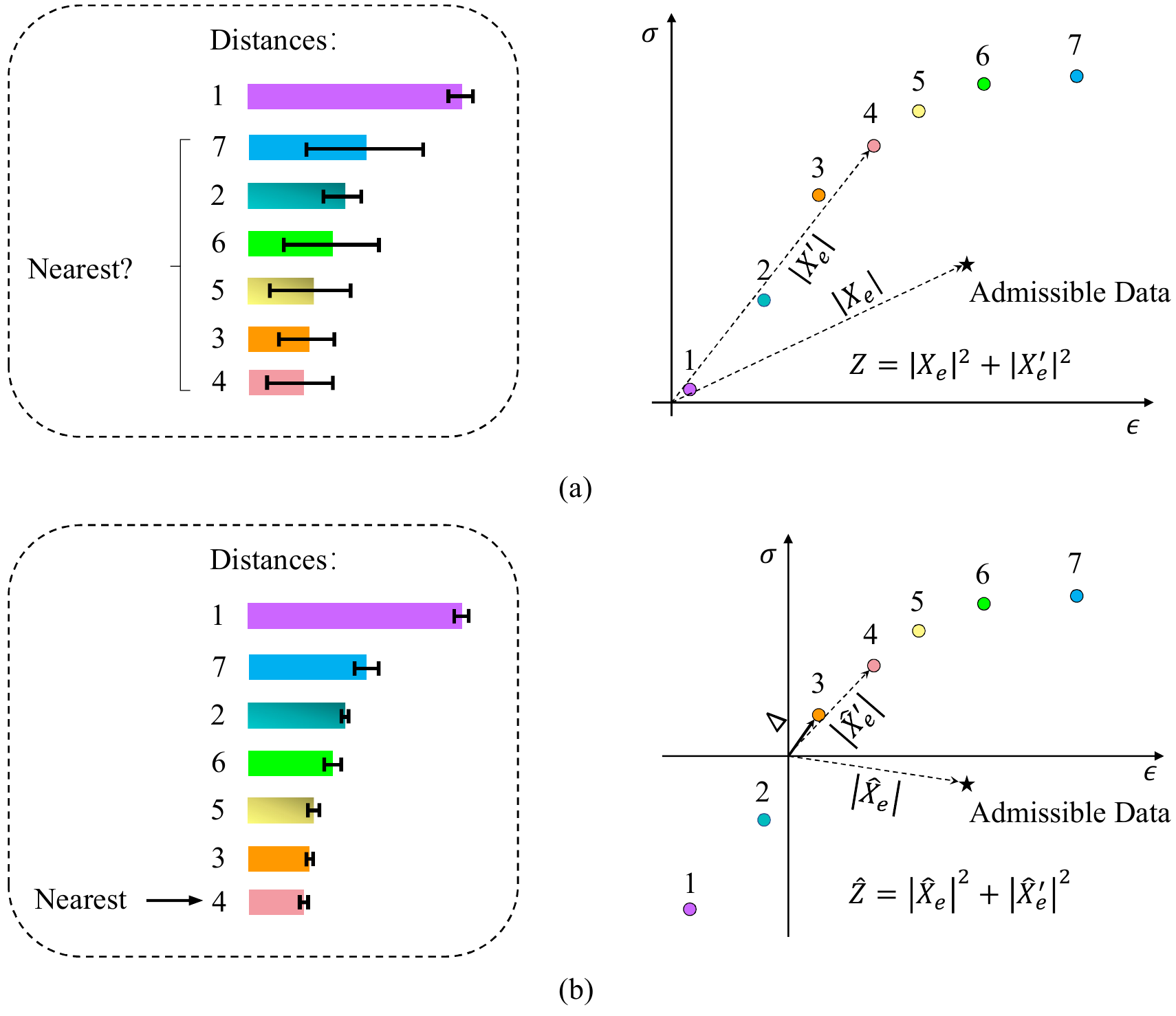}
\caption{Illustration of the adaptive strategy. (a) Due to the error in distance estimation, finding the correct nearest data is challenging. (b) However, the translation of data reduces the error, increasing the probability of finding the correct nearest data.}
\label{Adaptive_strategy}
\end{figure}

To illustrate the influence of the distance estimation error on nearest-neighbor search, we present an example in \cref{Adaptive_strategy} (a), where a nearest-neighbor search task with one admissible data and seven material data is shown. The true distances $d$ are depicted with colored bars on the left, and the associated black range intervals indicate the 25th to 75th percentile of the $\bar{d}$ distribution (interquartile range), indicating that the estimated $\bar{d}$ is likely to fall in the range with a high degree of probability, and the length of this range is influenced by the estimation error $\epsilon_{\bar{d}}$. In this example, the nearest data should be the one marked with `4' since the corresponding distance is minimum. However, since the range intervals intersect with each other, all data except for the one marked with `1' may be regarded as the nearest to the admissible data, which makes it hard to find the correct nearest data. Consequently, it is challenging for qDD to converge.

\subsubsection{Error reduction}\label{Reducing}
To reduce the error of estimated $\bar{d}$, a proper approach is needed to reduce  $Z=\left| \bm{X}_{e} \right|^2+\left| \bm{X}'_{e} \right|^2$. Here, we propose an adaptive strategy to reduce $Z$, which relies on pre-processing the database with translations. 

In one iteration of qDD computing, the optimal material data $\bm{X}^{*}_{e}$ computed from the last iteration is assigned at each integration point first. Then, admissible data $\bm{X}_{e}$ is computed by solving the linear system \cref{DD_linear}. The adaptive strategy requires the following translation:
\begin{subequations}\label{translation}
\begin{align}
& \hat{\bm{X}}_{e}=\bm{X}_{e}-\bm{X}^{*}_{e}+\bm{\Delta}\\
& \hat{\bm{X}}'_{e}=\bm{X}'_{e}-\bm{X}^{*}_{e}+\bm{\Delta}
\end{align}
\end{subequations}
where $\hat{\bm{X}}_{e}$ and $\hat{\bm{X}}'_{e}$ denote translated $\bm{X}_{e}$ and $\bm{X}'_{e}$, respectively.  
Since amplitude encoding can not represent a vector of zeros easily, a user-defined vector $\bm{\Delta}$ with a small norm is used to make sure that $|\hat{\bm{X}}'_{e}|>0$. Note that $|\hat{\bm{X}}_{e}-\hat{\bm{X}}'_{e}|^2=|\bm{X}_{e}-\bm{X}'_{e}|^2=d$, so the true distance is not changed. 
While the $Z$ after translation is:
\begin{equation}\label{translation1}
\hat{Z}=|\hat{\bm{X}}_{e}|^2+|\hat{\bm{X}}'_{e}|^2 \approx |\bm{X}_{e}-\bm{X}^{*}_{e}|^2+|\bm{X}'_{e}-\bm{X}^{*}_{e}|^2
\end{equation}
where the term $\bm{\Delta}$ is ignored since it has a small norm. When qDD is approaching convergence,  $\hat{Z}$ is smaller than $Z$ for the material data near the final optimal data. In this manner, the distance estimation error is reduced for these data, and the convergence of qDD will be improved. For instance, in \cref{Adaptive_strategy} (b), the material data marked with `3' is assumed to be the optimal data $\bm{X}^{*}_{e}$. The admissible data, along with the whole database, are translated so that data `3' is near the coordinate origin. Compared to \cref{Adaptive_strategy} (a), the width of the interquartile ranges is obviously reduced, making it more probable to find the true nearest data. In addition, we would like to mention that the main purpose of the adaptive strategy is to reduce the error for the material data near the optimal data. Furthermore, to release the estimation error of the material data far away from the optimal data, we may integrate the adaptive strategy with the sub-domain search technique \cite{kuang2023data} in future research. This approach is achieved by performing a nearest-neighbor search only in the material data set near the optimal data.

With the adaptive strategy, the two quantum states encoded into the quantum computer in \cref{states2} become:
\begin{subequations}\label{states_a}
\begin{align}
& |\phi\rangle=\frac{1}{\sqrt{\hat{Z}}}(| \hat{\bm{X}}_{e} ||0\rangle-| \hat{\bm{X}}'_{e} ||1\rangle)\\
& |\psi\rangle=\frac{1}{\sqrt{2}}(|0\rangle|\hat{\bm{X}}_{e}\rangle+|1\rangle|\hat{\bm{X}}'_{e}\rangle)
\end{align}
\end{subequations}
Note that before quantum computing, we need to pre-compute $\hat{Z}$ and the components in $|\phi\rangle$ and $|\psi\rangle$ with classical computing. 
%The two components in $|\phi\rangle$ are $\frac{1}{\sqrt{\hat{Z}}}| \hat{\bm{X}}_{e} |$ and $\frac{1}{\sqrt{\hat{Z}}}| \hat{\bm{X}}'_{e} |$, and the components in $|\psi\rangle$ are two normalized data $\frac{1}{\sqrt{2}| \hat{\bm{X}}_{e} |} \hat{\bm{X}}_{e}$ and $\frac{1}{\sqrt{2}| \hat{\bm{X}}'_{e} |} \hat{\bm{X}}'_{e}$.
%, where $\sqrt{2}$ is used to make sure the sum of the amplitudes of the quantum state equals to 1. 
The components related to the translated material data $\hat{\bm{X}}'_{e}$ can be computed offline in advance, and stored in the database named adaptive database denoted by $\hat{\mathscr{D}}_{j}$, where the data $\bm{X}^{*}_{e}$ used for translation is the $j$th data in the original database $\mathscr{D}$, $j=1,...,N$. It takes $O(N^2D)$ classical hardware resources to store $N$ adaptive databases $\hat{\mathscr{D}}_{j}$. As for the components related to the admissible data $\hat{\bm{X}}_{e}$, they can be computed online and only brings cost $O(D)$, which is independent of $N$. During the iterations of qDD, the nearest-neighbor search is actually performed in the adaptive database for each translated admissible data. The detailed procedures and computational cost will be introduced in the next section, and the effectiveness of the adaptive strategy in qDD is validated in \cref{Validation}.

%%%%%%%%%%%%%%%%%%%%%%%%%%%%%%%%%%%%%%%%%%%%%%%%%%%%
\subsection{The complexity of qDD}\label{complexity}

\begin{figure}[t]
\centering
\includegraphics[width=15cm]{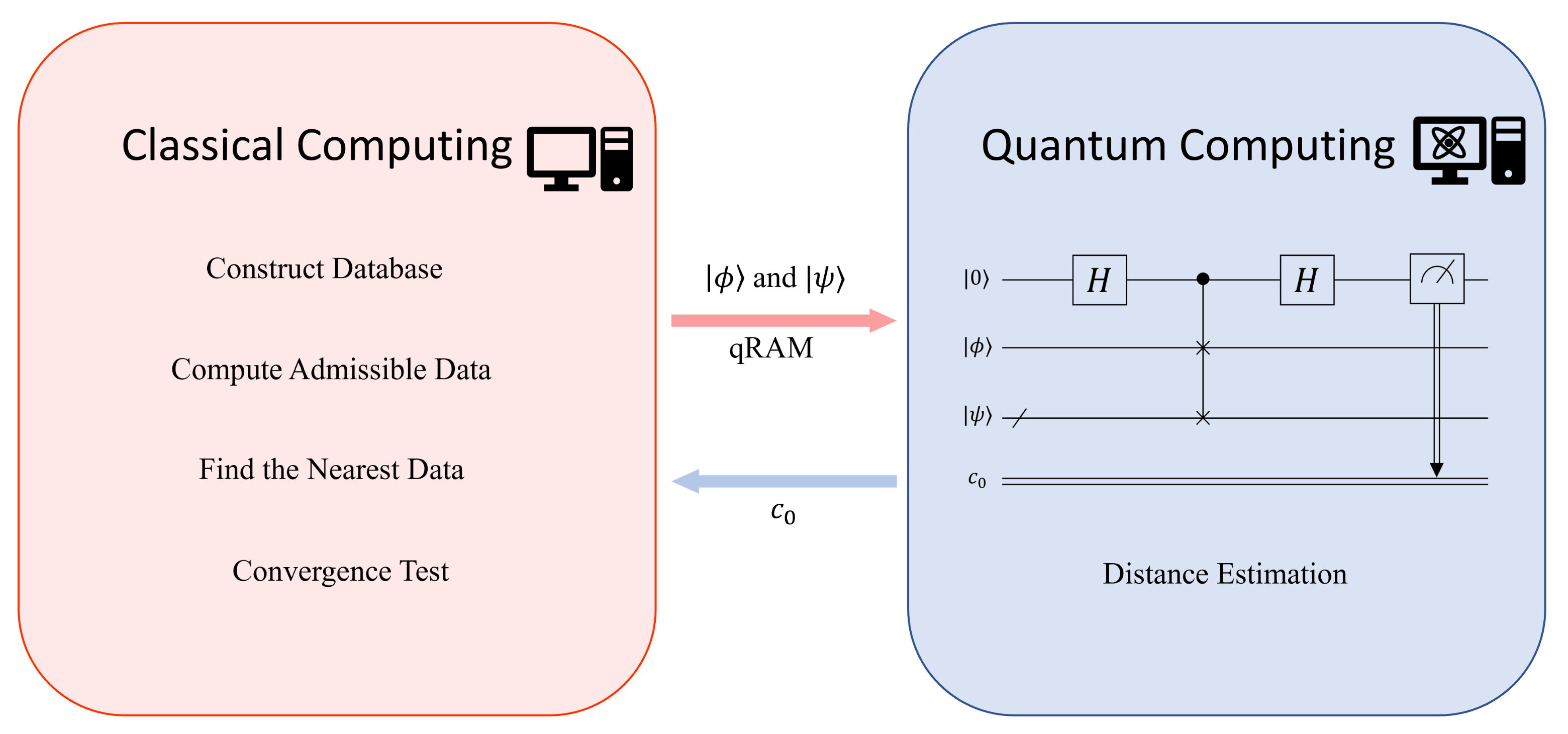}
\caption{The computational framework of qDD.}
\label{Framework}
\end{figure}

The framework of qDD is a hybrid of classical computing and quantum computing, as illustrated in \cref{Framework}. Quantum computing is used for distance estimation, and classical computing takes care of the rest of the procedures. A detailed flowchart of qDD is provided in \cref{flowchart}, and the distance estimation part is visually distinguished by a shaded box.
The complexity of distance estimation in qDD is presented in the following.

\begin{algorithm}

\small

\caption{qDD}
\begin{algorithmic}[0]
\Require Original database $\mathscr{D}$ containning $N$ data. Adaptive databases $\hat{\mathscr{D}}_{j}$, $j=1,...,N$. The number of repetitions $n_s$. A vector $\bm{\Delta}$ with a small norm. 
\State \textbf{Step 1: Initial assignments}
\State  Set number of iterations $k=0$.
\For{\(e = 1\) to \(M\)}
        \State Generate a random number $j_e^{(0)}\in \{1,...,N\}$. Set $\bm{X}^{*(0)}_{e}$ as $j_e^{(0)}$th data in $\mathscr{D}$.
\EndFor
\State \textbf{Step 2: Admissible data } 
\State Solve the linear problem in \cref{DD_linear}, and get admissible data $\bm{X}^{(k)}_{e}$ for $e=1,...,M$.
\State \textbf{Step 3: Nearest-neighbor search}
\For{\(e = 1\) to \(M\)}
					 \State $\hat{\bm{X}}^{(k)}_{e}=\bm{X}^{(k)}_{e}-\bm{X}^{*(k)}_{e}+\bm{\Delta}$.
					 \State Compute $|\hat{\bm{X}}^{(k)}_{e}|$ and $\frac{1}{\sqrt{2}|\hat{\bm{X}}^{(k)}_{e}|}\hat{\bm{X}}^{(k)}_{e}$.
					\For{\(i = 1\) to \(N\)}

					\State Read the $i$th data $|\hat{\bm{X}}'_{e}|$ and $\frac{1}{\sqrt{2}|\hat{\bm{X}}'_{e}|}\hat{\bm{X}}'_{e}$ in the adaptive database $\hat{\mathscr{D}}_{j_e^{(k)}}$.
					\State Compute $\hat{Z}=|\hat{\bm{X}}^{(k)}_{e}|^2+|\hat{\bm{X}}'_{e}|^2$, $\frac{1}{\sqrt{\hat{Z}}}|\hat{\bm{X}}^{(k)}_{e}|$, and $\frac{1}{\sqrt{\hat{Z}}}|\hat{\bm{X}}'_{e}|$.

      \begin{mdframed}[backgroundcolor=lightgray!20]
      \State \textbf{Distance estimation via the quantum algorithm } 
        \State Set the number of 0s in the measurments $V=0$.
        \For{\(t = 1\) to \(n_s\)}
        	\State Encode pre-computed $|\phi\rangle$ and $|\psi\rangle$ into quantum computer using qRAM. 
        	\State Execute the swap test in the quantum computer.
        	\State  Transfer the measurement result of the ancilla qubit in classical bit $c_0$.
        \If{$c_0=0$}
        \State $V\leftarrow V+1$
        \EndIf
        
        \EndFor
        \State Compute the distance ${{\bar d}_{i}} = \hat{Z}(4\frac{V}{{{n_s}}} - 2)$.
        \end{mdframed}

        \EndFor
        \State Sort the distances $\bar{d}_{i}$, where $i=1,...,N$. 
        \State Set $j_e^{(k+1)}\in \{1,..., N\}$ as the index corresponding to min($\bar{d}_{i}$). 
        \State Set $\bm{X}^{*(k+1)}_{e}$ as $j_e^{(k+1)}$th data in $\mathscr{D}$.
\EndFor
\State \textbf{Step 4: Convergence test}
        \If{$j_e^{(k+1)}=j_e^{(k)}$ for all $e=1,..., M$}
        \State \textbf{exit}.
        \Else
        \State $k \leftarrow k+1$, \textbf{goto} Step 2.
        \EndIf
\end{algorithmic}
\label{flowchart}
\end{algorithm}

\begin{table}[htbp]
  \centering
  \footnotesize
  \caption{The computational complexity of distance estimation in one nearest-neighbor search.}
    \begin{tabular}{cccccccc}
    \toprule
    \multicolumn{2}{c}{Procedures} & \multicolumn{2}{c}{Complexity} & \multicolumn{2}{c}{Repetitions} & \multicolumn{2}{c}{Total complexity} \\
    \midrule
    \multicolumn{2}{c}{\multirow{3}[6]{*}{Pre-compute $Z$, $|\phi\rangle$ and $|\psi\rangle$}} & $\hat{\bm{X}}_{e}=\bm{X}_{e}-\bm{X}^{*}_{e}+\bm{\Delta}$ & $O(D)$ & \multicolumn{2}{c}{\multirow{3}[6]{*}{1}} & \multicolumn{2}{c}{\multirow{5}[10]{*}{$O(N\log D)$}} \\
\cmidrule{3-4}    \multicolumn{2}{c}{} & $| \hat{\bm{X}}_{e} |$ & $O(D)$ & \multicolumn{2}{c}{} & \multicolumn{2}{c}{} \\
\cmidrule{3-4}    \multicolumn{2}{c}{} & $\frac{1}{\sqrt{2}| \hat{\bm{X}}_{e} |} \hat{\bm{X}}_{e}$ & $O(D)$ & \multicolumn{2}{c}{} & \multicolumn{2}{c}{} \\
\cmidrule{1-6}    \multicolumn{2}{c}{\multirow{2}[4]{*}{Distance estimation}} & Encoding via qRAM & $O(\log D)$ & \multicolumn{2}{c}{\multirow{2}[4]{*}{$N$}} & \multicolumn{2}{c}{} \\
\cmidrule{3-4}    \multicolumn{2}{c}{} & Swap test & $O(1)$ & \multicolumn{2}{c}{} & \multicolumn{2}{c}{} \\
    \bottomrule
    \end{tabular}%
  \label{table1}%
\end{table}%

\cref{table1} shows the computational complexity of distance estimation in one nearest-neighbor search. First, we need to pre-compute $Z$ and the components in $|\phi\rangle$ and $|\psi\rangle$ with classical computing. The terms related to the admissible data require $O(D)$ cost of online computing, which only needs to be performed once. As for the terms related to the material data, they can be directly read from the adaptive databases. Then, encoding the quantum states into the quantum computer via qRAM requires $O(\log D)$ costs. Next, the swap test costs only $O(1)$ since only 3 gates are used, which is independent of the dimension $D$. Therefore, one distance estimation requires $O(\log D)$ costs, which need to be executed $N$ times to find the nearest neighbor, with $N$ being the number of material data in the database. Consequently, the total computational cost is $O(D)+O(N\log D)$. 
%Typically, the number of data $N$ is much greater than the dimension $D$, resulting in a simplified complexity of $O(N\log D)$.
This computational complexity can be simplified to $O(N\log D)$, considering that the number of data $N$ is usually much larger than the data dimension $D$.

In a word, with the help of the quantum algorithm for distance estimation, the computational complexity is reduced from $O(ND)$ to $O(N \log D)$, an exponential acceleration in the term of dimension $D$.

%
%
%%%%%%%%%%%%%%%%%%%%%%%%%%%%%%%%%%%%%%%%%%%%%%%%%%%%
%%%               Validation                    %%%
%%%%%%%%%%%%%%%%%%%%%%%%%%%%%%%%%%%%%%%%%%%%%%%%%%%
 \section{Validation}\label{Validation}

In this section, a simple one-dimensional stretched bar is investigated to verify the convergence and accuracy of qDD, and to show the effectiveness of the adaptive strategy. 
Due to the limitation on the development of hardware level of quantum computers \cite{preskill2018quantum,biamonte2017quantum}, the open-source quantum computer simulator platform Qiskit from IBM \cite{Qiskit} is used. It allows one to run the quantum algorithm with quantum computer simulation methods on a classical computer.

\begin{figure}[!hbtp]
\centering
\includegraphics[width=7cm]{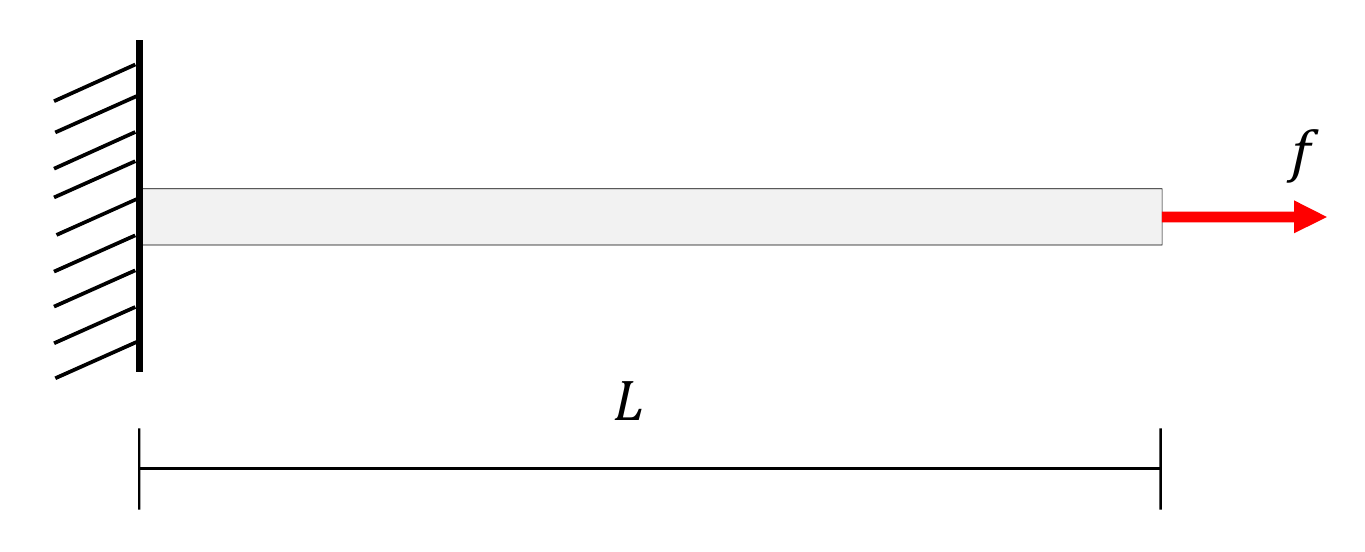}
\caption{A one-dimensional bar is used as an example to validate qDD.}
\label{1Dbar}
\end{figure}

As shown in \cref{1Dbar}, the length of the bar $L$ is 100 mm, and its cross-section area is 1 mm$^2$. A force $f=50$ N is applied on the right. The bar is discretized with 10 bar elements of the same length. 
The material behavior is assumed to obey the Ramberg-Osgood relation, which is used to construct the material database:
\begin{equation}
	\mathscr{D}=\left\{(\sigma,\epsilon)\bigg|\epsilon=\frac{\sigma}{E}+{\alpha\frac{\sigma}{E}\left(\frac{\left|\sigma\right|}{\sigma_0}\right)^{n-1}}\right\}
\end{equation}
where Young's modulus $E=10^4$ MPa, the yield offset $\alpha=3$, the yield stress $\sigma_0=10$ MPa and the hardening exponent for plasticity $n=3$. The data are sampled in the range $\sigma \in [-5, 60]$ MPa, which is uniformly distributed with 100 points.

\begin{figure}[!hbtp]
\centering
\includegraphics[width=9.7cm]{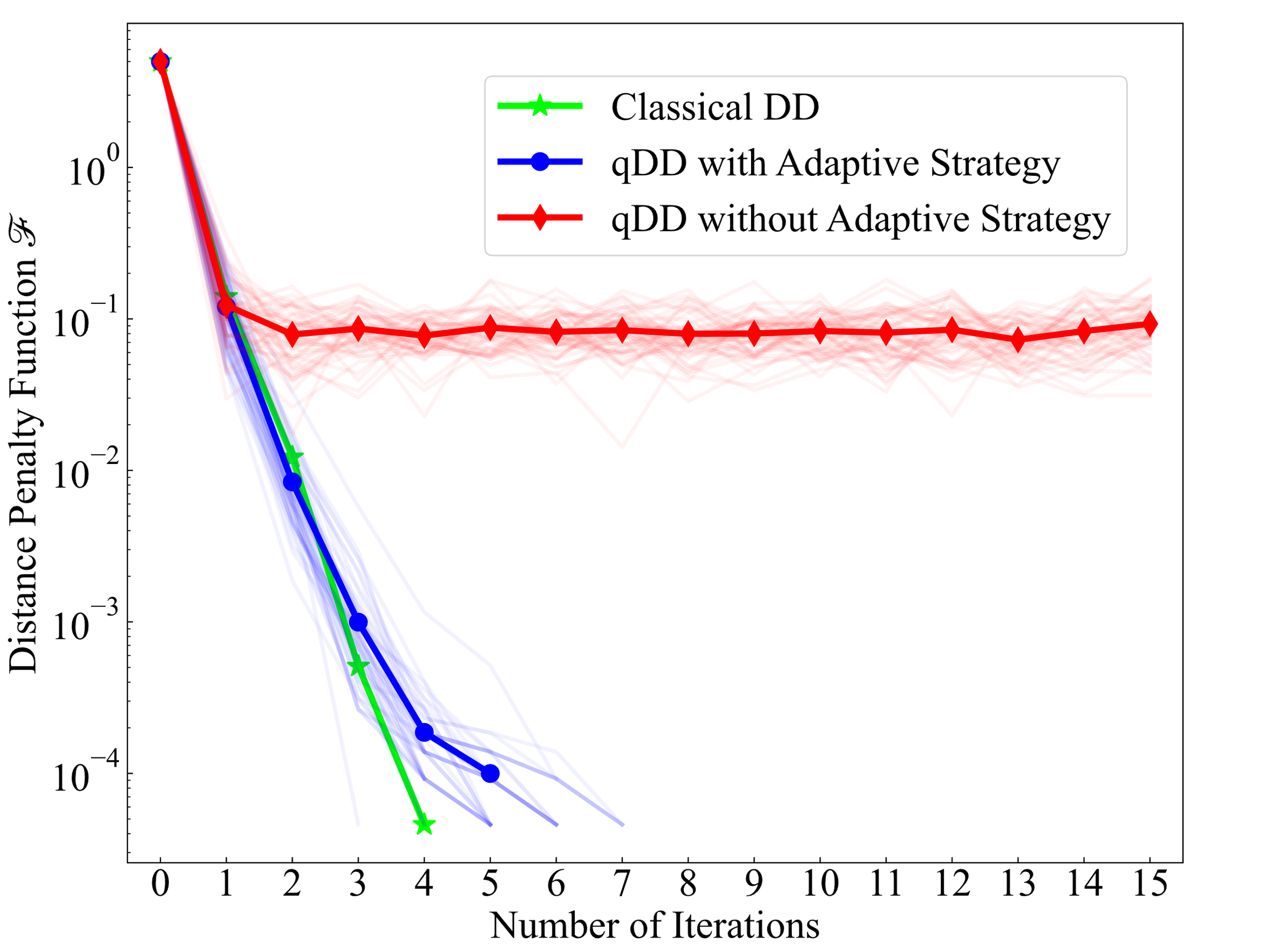}
\caption{The distance penalty function $\mathscr{F}$ versus the number of iterations.}
\label{F_fig}
\end{figure}

\begin{figure}[!hbtp]
\centering
\includegraphics[width=9cm]{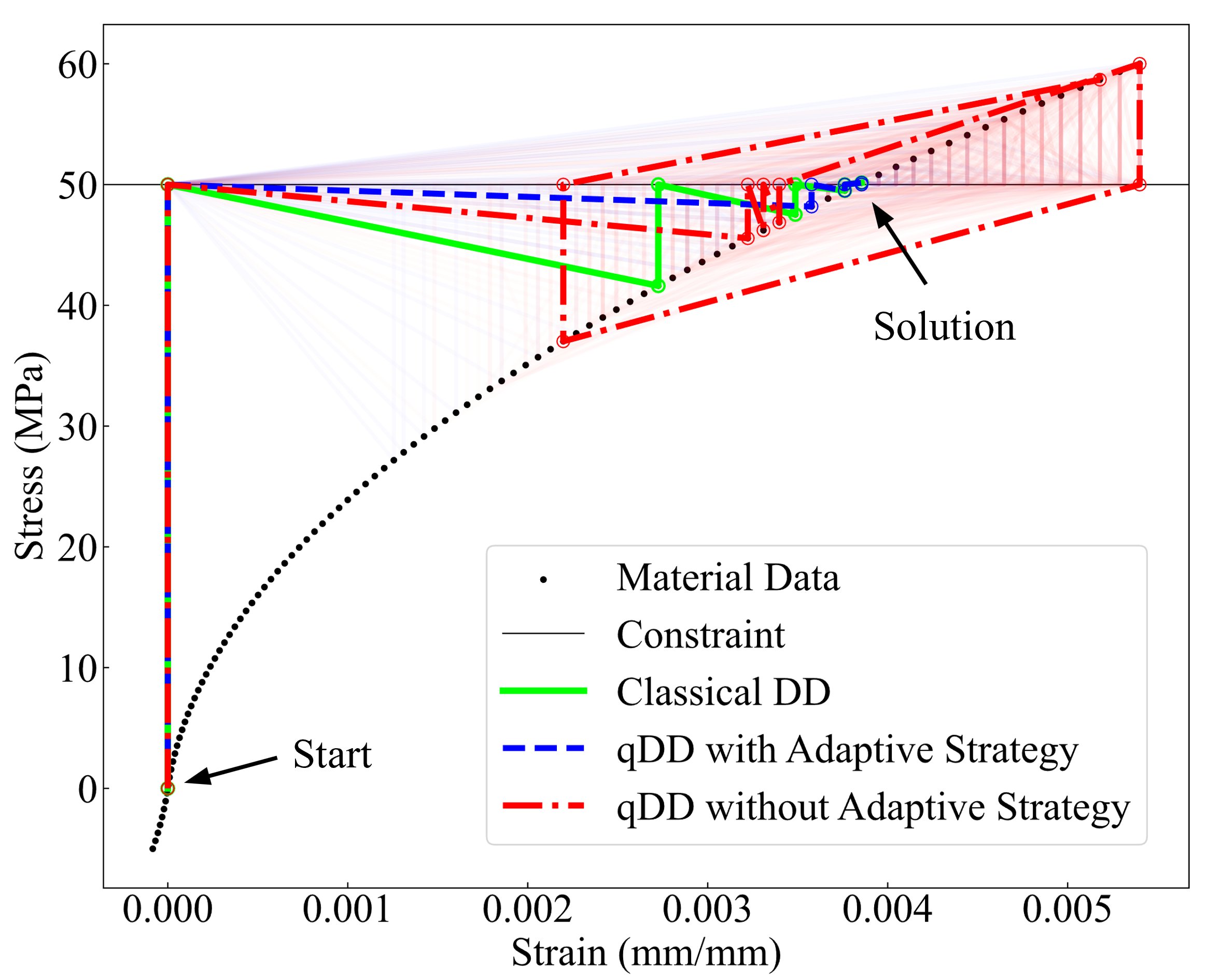}
\caption{The evolution curves of $(\sigma_e, \epsilon_e)$ at the right-most integration point.}
\label{iter}
\end{figure}

To demonstrate the iteration process of the data-driven computing, we define a distance penalty function $\mathscr{F}$ \cite{Ortiz2016} as the sum of the distances at all integration points $\sum_{e=1}^{m}{w_e}\mathscr{F}_e(\bm{z}_e,\bm{z}_e^*)$ shown in \cref{Eq01}.  \cref{F_fig} shows the evolution of the distance penalty function $\mathscr{F}$ versus the number of iterations obtained by three methods: classical DD, qDD with the adaptive strategy, and qDD without the adaptive strategy. The initial material data for each integration point is zero. The number of repetitions $n_s$ for the two qDD methods is set to be 180.
Due to the probabilistic characteristics of the quantum algorithm,  results with slight differences are obtained if one repeated the calculation with qDD. Thus, the averaged results of 40 repeated simulations are adopted.  
The qDD with adaptive strategy consumes less than 7 iterations to achieve convergence, resulting in a  distance penalty function $\mathscr{F}$ similar to that obtained by the classical DD. However, the qDD without the adaptive strategy results in an obvious erroneous solution where the distance penalty function $\mathscr{F}$ is three orders of magnitude larger than the classical DD. 
\cref{iter} also shows the evolution curves of $(\sigma_e, \epsilon_e)$ at the right-most integration point for the three methods.  Due to the error of distance estimation in the qDD without the adaptive strategy, this method fails to find the correct nearest data, and thus fails to obtain converged results. To clarify, only the first 8 iterations of this method are shown.
In comparison, all evolution curves of qDD with the adaptive strategy converge to certain results which are close to the reference result of classical DD.
In total, the qDD with adaptive strategy can significantly reduce the error of  distance estimation and result in reliable solutions. It will be used in the following examples.

\begin{figure}[!hbtp]
\centering
\includegraphics[width=10cm]{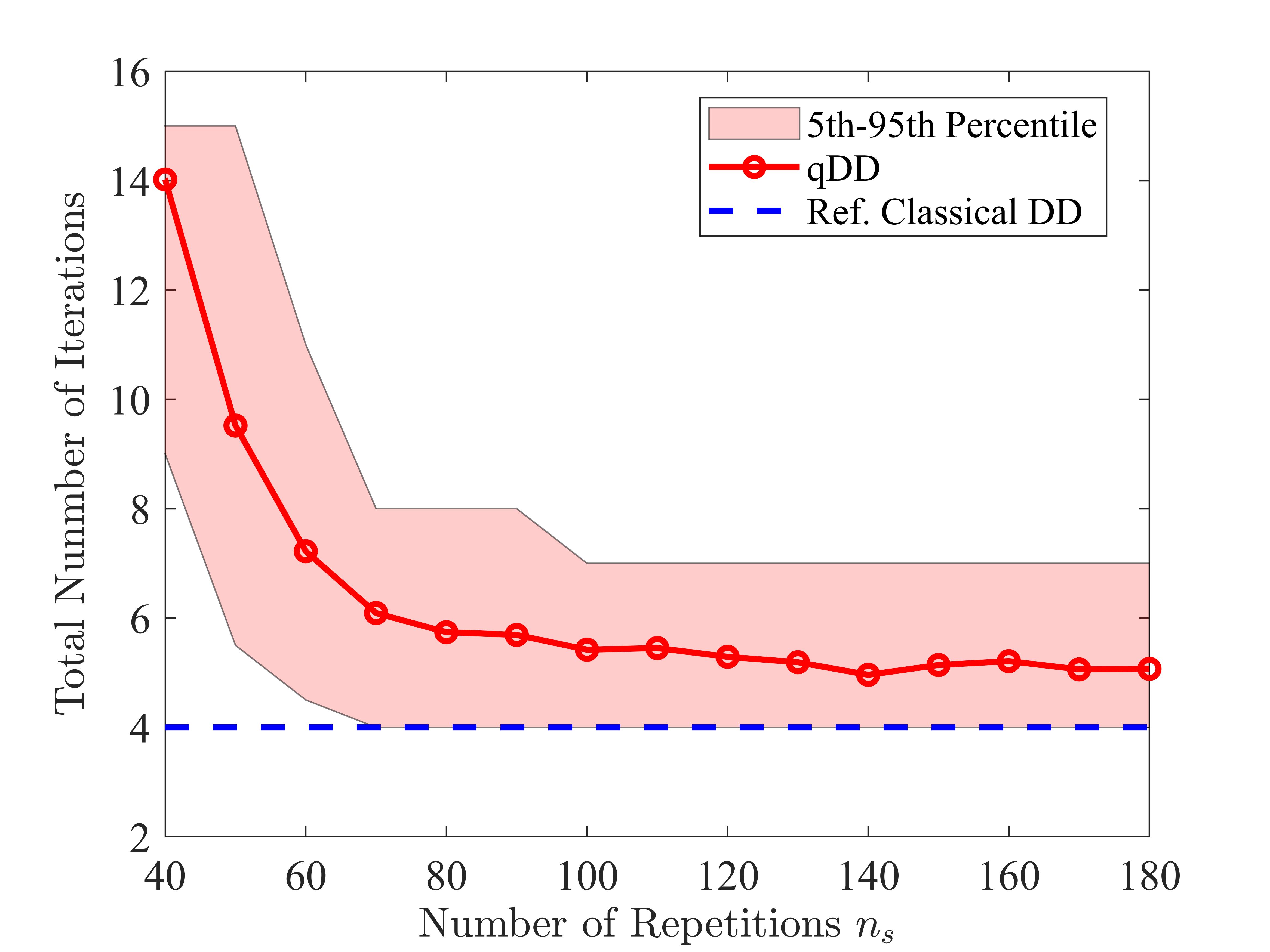}
\caption{The total number of iterations versus the number of repetitions $n_s$.}
\label{Iter_ns}
\end{figure}

 \begin{figure}[!hbtp]
\centering
\includegraphics[width=10cm]{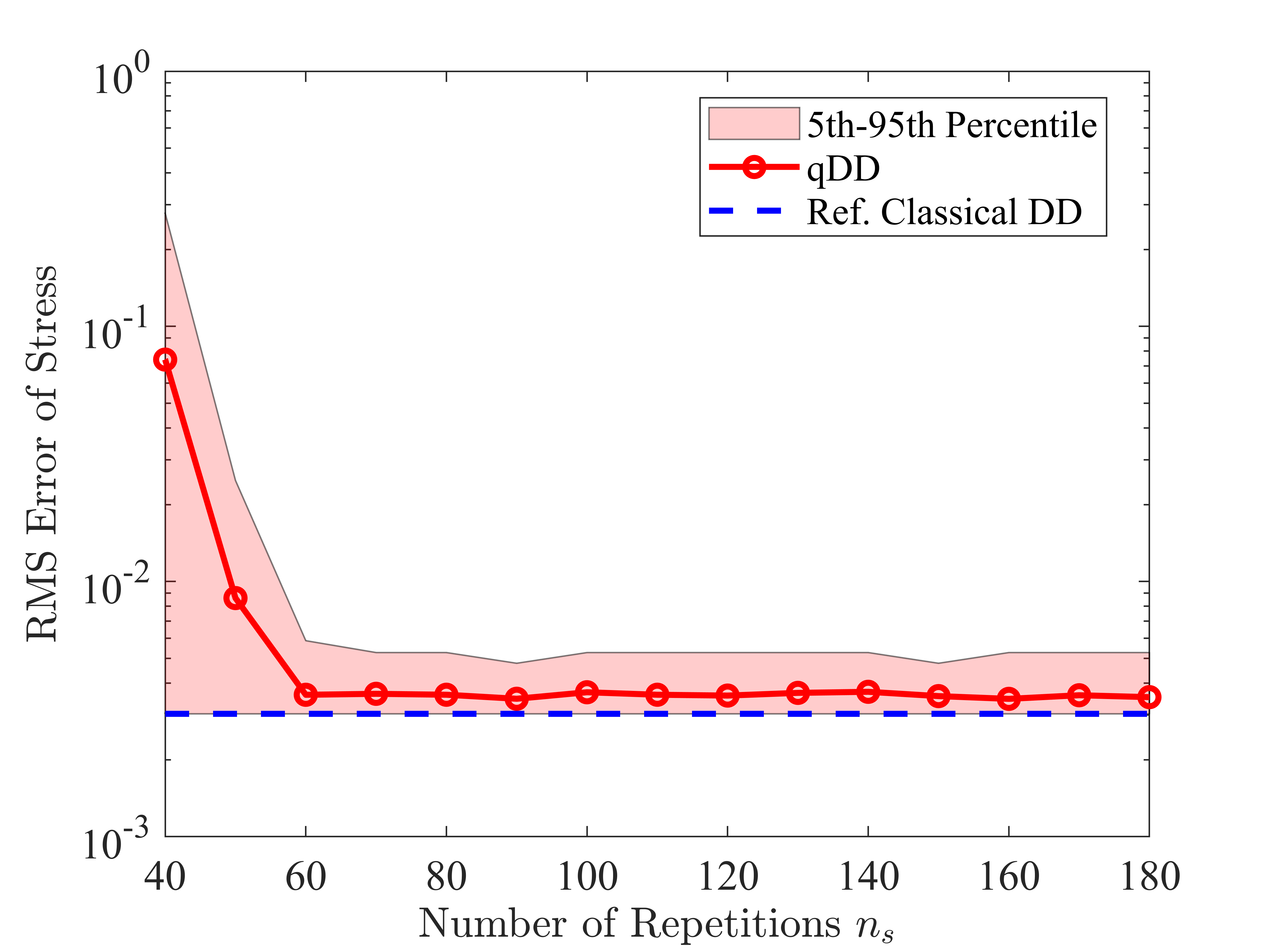}
\caption{The RMS error of stress versus the number of repetitions $n_s$.}
\label{RMS_ns}
\end{figure}

\cref{Iter_ns} presents the total number of iterations versus the number of repetitions $n_s$, whereas \cref{RMS_ns}  presents the RMS error of stress for qDD with adaptive strategy versus $n_s$. 
The results of classical DD marked with the blue dashed line are taken as reference solutions.  We perform 100 times of simulations with the same $n_s$ to demonstrate the statistical characteristics of qDD. The average values are shown in red lines, and the 5th to 95th percentile of the distributions are shown in the light red band.
When increasing the number of repetitions to about 100, both the number of iterations and the RMS error of stress decrease rapidly. When the number of repetitions exceeds 100, these two parameters remain nearly unchanged, and are almost the same as that of classical DD. This is consistent with the error analysis \cref{Var_d}. When the $n_s$ is large enough,  the same distance estimation error as classical computing can be obtained, and the number of iterations and RMS error would be the same as classical DD.

%
%
%%%%%%%%%%%%%%%%%%%%%%%%%%%%%%%%%%%%%%%%%%%%%%%%%%%%
%%%               Real                    %%%
%%%%%%%%%%%%%%%%%%%%%%%%%%%%%%%%%%%%%%%%%%%%%%%%%%%
\section{Experiment of qDD on a real quantum computer}\label{real}

The proposed qDD is further tested on a superconducting quantum computer called WuYuan from OriginQ \cite{Origin}, which offers 6-qubit quantum computing sources. Here, we test qDD with a simple spring-bar structure \cite{ortiz2017data}. 

\begin{figure}[!hbtp]
\centering
\includegraphics[width=8cm]{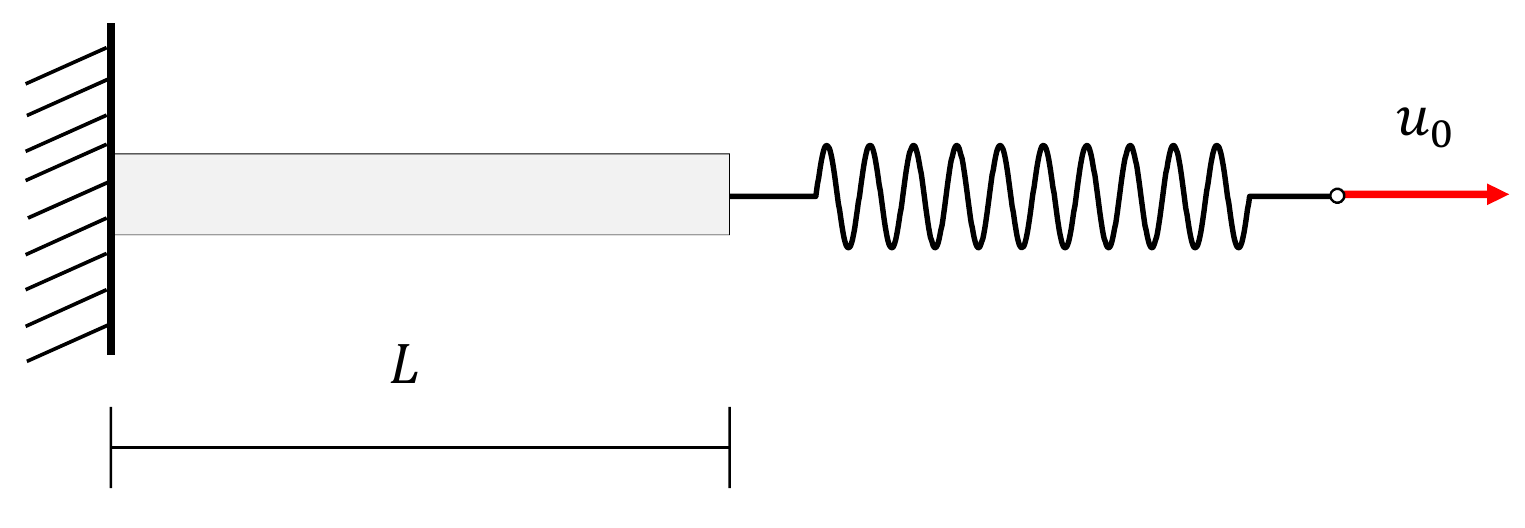}
\caption{The spring-bar structure.}
\label{Spring}
\end{figure}

As shown in \cref{Spring}, the spring is stretched with the displacement $u_0=0.5$ mm, and the spring constant $k$ is 140 N/mm.  The length of the bar $L$ is set to be 100 mm, and the cross-section area is 1 mm$^2$.  The material database of the bar is generated using the same parameters in \cref{Validation}, and 20 data are uniformly sampled in the range $\sigma \in [-5, 60]$ MPa.

\begin{figure}[!hbtp]
\centering
\includegraphics[width=\textwidth]{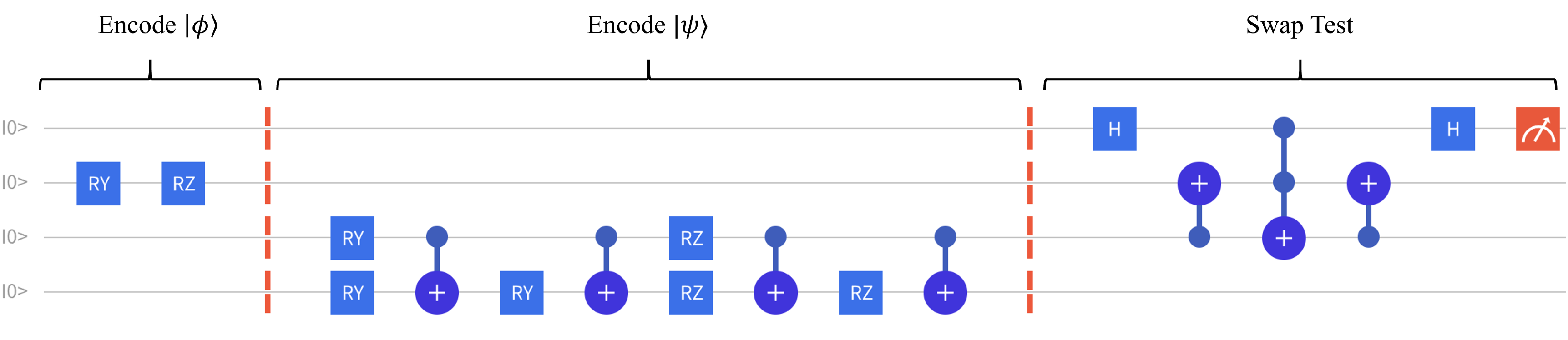}
\caption{The quantum circuit for distance estimation on the OriginQ quantum computer. The first two parts are used for encoding $|\phi\rangle$ and $|\psi\rangle$ into the quantum computer \cite{mottonen2004transformation}. The third part is the swap test. }
\label{qDD_real}
\end{figure}

Further development is still needed before qRAM is put into practical use for a real quantum computer, thus a state preparation method proposed by Mottonen et al. \cite{mottonen2004transformation} is used to encode the quantum states. The main idea is to encode the two quantum states $|\phi\rangle$ and $|\psi\rangle$ through a series of unitary transformations. These transformations require one-qubit elementary rotation gates and CNOT gates, and the information of the two quantum states is encoded through the parameters of the gates. Note that the complexity of this method is $O(D)$ \cite{mottonen2004transformation} rather than $O(\log D)$ by using qRAM. \cref{qDD_real} shows the quantum circuit performed on the OriginQ quantum computer, where four qubits are required for the spring-bar problem, and the circuit mainly consists of three parts.  The first part consists of an RY gate and an RZ gate to encode $|\phi\rangle$ into the second qubit. The second part consists of a series of one-qubit elementary rotation gates and four CNOT gates, and it is used to encode $|\psi\rangle$ into the third and fourth qubits. The third part is the swap test, where the controlled-SWAP gate is decomposed into two CNOT gates and one Toffoli gate.

\begin{figure}[!hbtp]
\centering
\includegraphics[width=12cm]{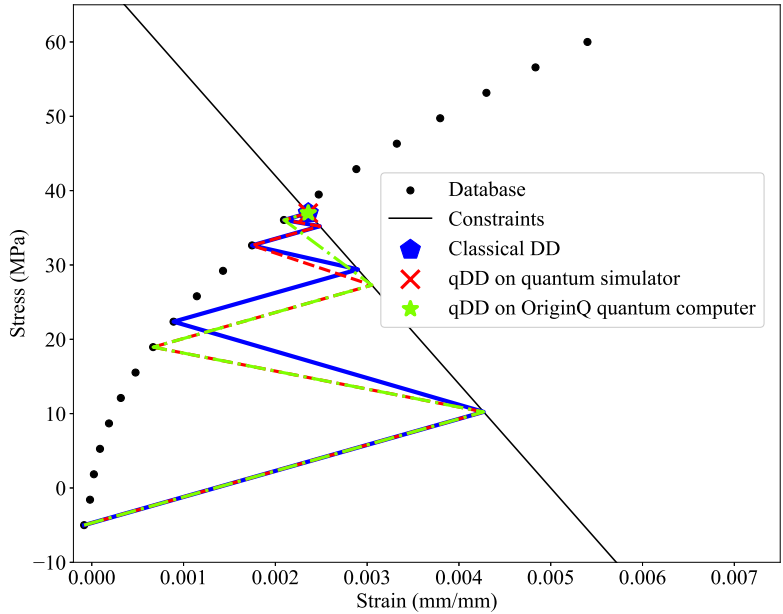}
\caption{The evolution curves of qDD on the simulator and the OriginQ quantum computer.}
\label{real_dis}
\end{figure}

\cref{real_dis} shows the evolution curves of  $(\sigma_e, \epsilon_e)$ in the bar, which are obtained by the qDD on the OrignQ quantum computer, the classical DD, and qDD on the simulator. To ensure the computational accuracy, the number of repetitions of running the circuit $n_s$ is set to be 10000. The convergent results agree very well for the three methods, which further demonstrates the potential usage of the proposed qDD.
The hardware noise in a real quantum computer can not be neglected, especially when the database is large. Further large-scale computing and testing need to be conducted if computing sources become more abundant and easier to use in the future.

%
%
%%%%%%%%%%%%%%%%%%%%%%%%%%%%%%%%%%%%%%%%%%%%%%%%%%%%
%%%               Test                    %%%
%%%%%%%%%%%%%%%%%%%%%%%%%%%%%%%%%%%%%%%%%%%%%%%%%%%
\section{Numerical test for two-dimensional example}\label{plate}

\begin{figure}[!hbtp]
\centering
\includegraphics[width=10cm]{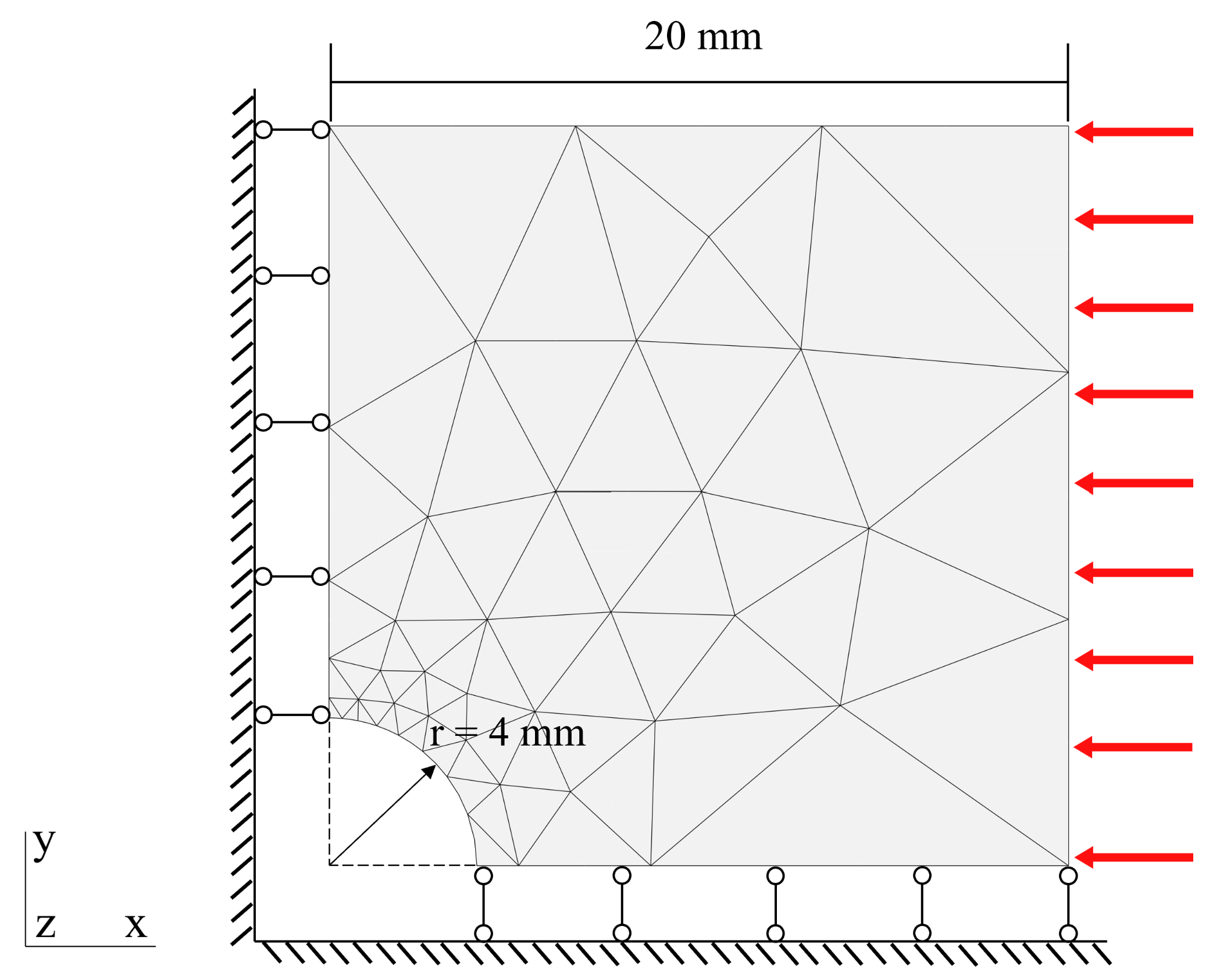}
\caption{A plate with a hole for the numerical test.}
\label{Plate}
\end{figure}

In this section, we present a numerical test of qDD with a two-dimensional example on the quantum computer simulator Qiskit, and the configuration of the example is shown in \cref{Plate}. The length of the square plate is 20 mm, and there is a quarter of a hole in the left-bottom corner with a radius of 4 mm. The left edge is fixed along the $x$ direction, and the bottom edge is fixed along the $y$ direction. The right edge is applied with a pressure of 400 MPa. The configuration is discretized with 6-node triangle finite elements.

\begin{figure}[t]
\centering
\includegraphics[width=15cm]{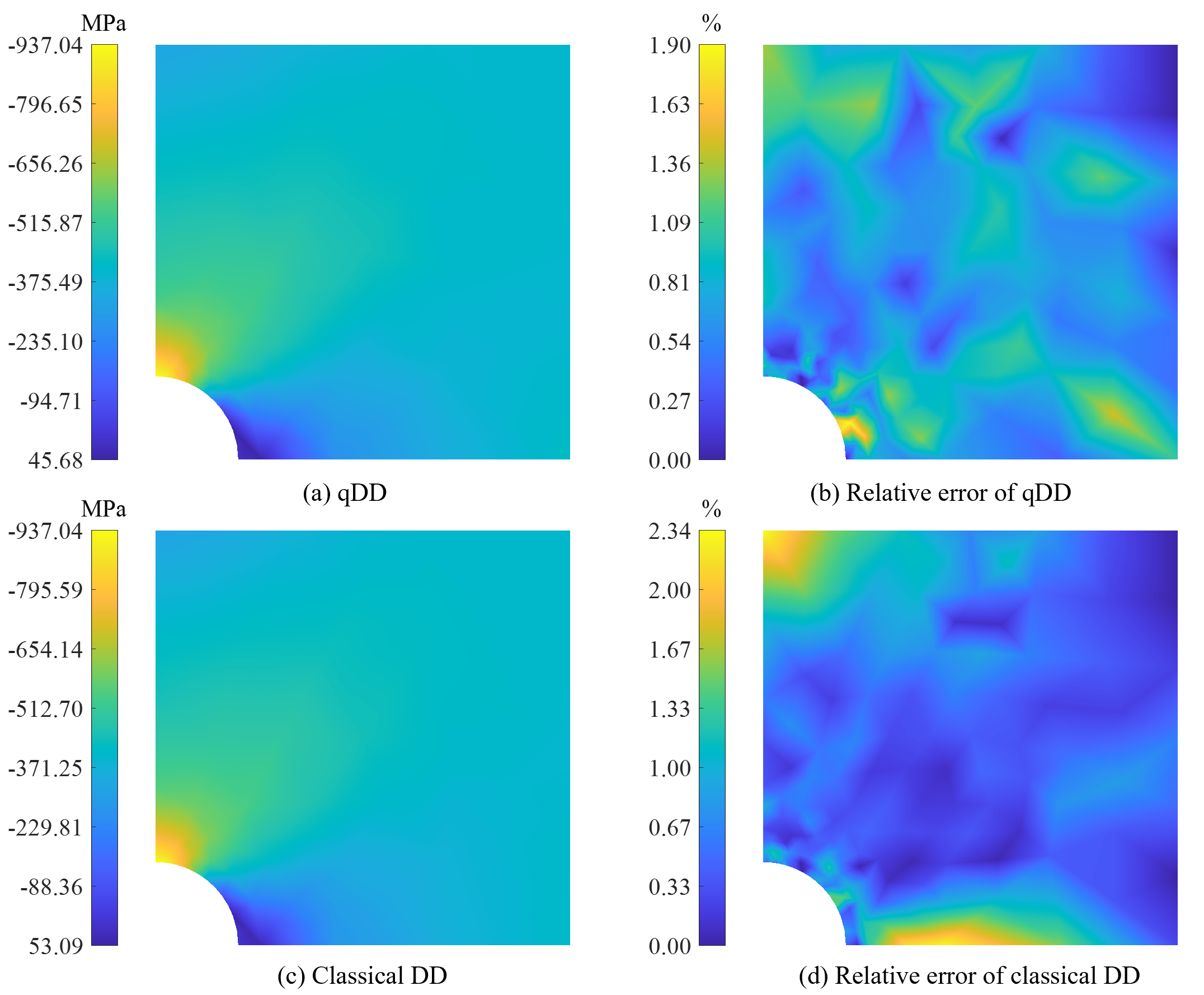}
\caption{Distribution of $\sigma_x$ and its relative errors via qDD and classical DD, where the solution via classical FEM is used as the reference.}
\label{Stress}
\end{figure}

Again, the Ramberg-Osgood material model is used to generate the material database:
\begin{equation}\label{Ramberg_Osgood}
	\mathscr{D}=\left\{(\bm{\sigma},\bm{\epsilon})\bigg|	E\bm{\epsilon}=(1+\nu)\bm{\sigma}^d-(1-2\nu)p\textbf{I}+\dfrac{3}{2}\alpha\left(\dfrac{\bar{\sigma}}{\sigma_0}\right)^{n-1}\bm{\sigma}^{d}\right\}
\end{equation}
where Young's modulus $E={10}^5$ MPa, Poisson's ratio $\nu=0.3$, the yield offset $\alpha=0.2$, the yield stress $\sigma_0=300$ MPa, the hardening exponent for plasticity $n=3$. The equivalent hydrostatic stress $p=-\frac{1}{3}\bm{\sigma}:\textbf{I}$, the stress deviator $\bm{\sigma}^d=\bm{\sigma}+p\textbf{I}$, the Mises equivalent stress $\bar{\sigma}=\sqrt{\frac{3}{2}\bm{\sigma}^d:\bm{\sigma}^d}$ and $\textbf{I}$ is an unit tensor. For plane stress problem, three stress components $\bm{\sigma}=[\sigma_x,\sigma_y,\sigma_{xy}]$ are considered. To generate the database, we uniformly sample the stress in ranges: $\sigma_x \in [-1000, 10]$ MPa, $\sigma_y \in [-250, 600]$ MPa, and $\sigma_{xy} \in [-150, 350]$ MPa to generate $100^3$ data.

For qDD, the number of repetitions $n_s$ is set to be 5000. As references, classical FEM and classical DD are used to solve the same problem. The simulation results are shown in \cref{Stress}, where the distribution of $\sigma_x$ and its relative errors via qDD and classical DD are shown. The maximum relative errors of qDD and classical DD are both  around 2\%, and the accuracy can be improved with a high-density database. The iteration numbers of qDD and classical DD are 21 and 9. According to the validation results in \cref{Validation}, the iteration number of qDD can be further reduced with a greater number of repetitions $n_s$. In addition, classical DD takes 0.1 hours to converge, whereas qDD requires approximately 69 hours. This difference arises from the inherent computational expense of simulating a quantum system using a classical computer, as it necessitates extensive matrix manipulations on the quantum state vector \cite{jones2019quest}. Nevertheless, the results show that qDD is able to achieve a similar accuracy as classical DD.

%
%
%%%%%%%%%%%%%%%%%%%%%%%%%%%%%%%%%%%%%%%%%%%%%%%%%%%%%%%%%%%%%%%%%%%%%%%%%%%
%%%                          Discussion                                 %%% 
%%%%%%%%%%%%%%%%%%%%%%%%%%%%%%%%%%%%%%%%%%%%%%%%%%%%%%%%%%%%%%%%%%%%%%%%%%%
\section{Conclusion}\label{Discussion}

This paper introduces a novel quantum computing enhanced data-driven computational framework that leverages a quantum algorithm for distance calculation. Taking advantage of quantum computing, the computational complexity of distance calculation is reduced from $O(D)$ to $O(\log D)$, where $D$ represents the dimension of the material data.
The validation example on the simulator demonstrates the convergence and accuracy of qDD, as qDD achieves similar accuracy to classical DD with a similar number of iterations.
Moreover, we test qDD on a real quantum computer provided by OriginQ,
and qDD successfully converges to the solution of classical DD. 
Finally, we present a numerical example of qDD in a two-dimensional problem on the simulator to illustrate its accuracy and potential.

While the proposed qDD is promising, it still comes with certain shortcomings: 
(1) It relies a lot on fault-tolerant quantum computers. Considering the current era of NISQ devices \cite{preskill2018quantum}, the error rates in these quantum computers make it challenging to solve practical problems effectively. 
(2) Though proof-of-principle demonstrations of qRAM have been performed, encoding classical data into the quantum computer 
is still a technological problem \cite{biamonte2017quantum, aaronson2015read}. 
(3) 
The adaptive strategy proposed in qDD requires significant storage resources, to be more specific, $O(N^2)$ storage for the adaptive database.
% The adaptive strategy proposed in qDD requires significant storage resources. Specifically, it demands $O(N^2)$ storage for the adaptive database. 
This requirement could be  impractical or even impossible to implement when dealing with large values of $N$. 
(4) Most of the procedures of qDD are still performed on the classical computer, these procedures may also be integrated with quantum computing in future research \cite{montanaro2016quantum,raisuddin2022feqa}.
Nevertheless, the introduction of qDD demonstrates the potential of quantum computing in data-driven computational mechanics, and it explores a new computational paradigm for further advancements in the field.

%(4) Although reducing computational complexity from $O(D)$ to $O(\log D)$ is a notable improvement, it may not be as impressive as the reduction achieved by the $k$-d tree data structure. However, $k$-d tree data structure may be naturally combined with the quantum algorithm for distance estimation, and achieves acceleration in both the dimension $D$ and the number of data $N$, resulting in a complexity of $O(\log D \log N)$.

%
%
%%%%%%%%%%%%%%%%%%%%%%%%%%%%%%%%%%%%%%%%%%%%%%%%%%%%
%%%                ACKNOWLEDGEMENTS              %%%
%%%%%%%%%%%%%%%%%%%%%%%%%%%%%%%%%%%%%%%%%%%%%%%%%%%%
\section*{Acknowledgements}

This work has been supported by the National Key R\&D Program of China (Grant No. 2022YFE0113100) and the National Natural Science Foundation of China (Grant Nos. 11920101002, 12202322 and 12172262).

%
%
%%%%%%%%%%%%%%%%%%%%%%%%%%%%%%%%%%%%%%%%%%%%%%%%%%%%
%%%                  BIBLIOGRAPHY                %%%
%%%%%%%%%%%%%%%%%%%%%%%%%%%%%%%%%%%%%%%%%%%%%%%%%%%%
\bibliographystyle{elsarticle-num}
\bibliography{QDD}
\end{document}